\documentclass[twocolumn,showpacs,prl,aps,amsmath,amssymb]{revtex4}
\usepackage{mathrsfs,epsfig}
\def\<{\langle}\def\>{\rangle}
\def\set#1{{\sf #1}}
\def\Rng{\set{Rng}}\def\Ker{\set{Ker}}\def\Supp{\set{Supp}}\def\Klm{{\mathscr X}}
\def\sH{\set{H}}\def\sK{\set{K}}
\def\Proof{\medskip\par\noindent{\bf Proof. }}\def\qed{\medskip\par}
\def\N#1{\left|\!\left|#1\right|\!\right|}\def\n#1{|\!|#1|\!|}\def\eg{e. g. }
\def\Proof{\medskip\par\noindent{\bf Proof. }}
\newtheorem{lemma}{Lemma}
\newtheorem{corollary}{Corollary}

\begin{document}
\title{There exist non orthogonal quantum measurements that are
perfectly repeatable}
\author{F. Buscemi, G. M.  D'Ariano and P. Perinotti}
\affiliation{QUIT Group, Unit\`a INFM and Dipartimento di Fisica ``A. Volta'', Universit\`a di Pavia, via A.
  Bassi 6, I-27100 Pavia, Italy} \homepage{http://www.qubit.it}
\date{\today}
\begin{abstract}
We show that, contrarily to the widespread belief, in quantum
mechanics repeatable measurements are not necessarily described by
orthogonal projectors---the customary paradigm of {\em observable}. 
Nonorthogonal repeatability, however, occurs only for infinite
dimensions. We also show that when a non orthogonal repeatable 
measurement is performed, the measured system retains some
``memory" of the number of times that the measurement has been performed.  
\end{abstract}
\maketitle
The main paradigm of quantum mechanics is the unavoidable
disturbance of the measurement on the measured system. This has
obvious disruptive consequences for an objective interpretation\cite{Mittelstaedt} of the
physical experiment. In order to retain some 
objectivity, one supposes the feasibility of some ``canonical"
measurements that can be actually regarded as the process of
``seizing" a property/quantity possessed by the system independently of
the measurement, thus assuming the existence of perfect 
measurements that satisfy the {\em repeatability
hypothesis} formulated by von Neumann\cite{vonneu}: if  
a physical quantity is measured twice in succession in a system, then
we get the same value each time. From this hypothesis it is then
concluded that the state after the measurement is the eigenvector
corresponding to the measurement outcome as the eigenvalue. 
In the conventional approach to quantum measurements,
this is the content of the so-called  von Neumann ``collapse"
postulate, which von Neumann posed as a kind of  
universal law, based on the Compton and Simmons experiment. 
\par Actually, as von Neumann himself admitted, for a degenerate
observable there are many different ways of satisfying the
repeatability hypothesis, with the state after the measurement given
by any mixture of eigenstates corresponding to the same outcome. The
concept of degenerate observable is crucial at foundational level
(i. e. to define local measurements on many particles), and in order
to retain repeatability, further physical hypotheses are needed 
to characterize a ``canonical" measurement. Such additional hypothesis
was introduced by 
L\H{u}ders\cite{Lud51} in form of a requirement of {\em least
disturbance}, leading to the von Neumann-L\H{u}ders {\em projection
postulate}, according to which the measurement of a discrete observable 
projects the state orthogonally on the eigenspace corresponding to the
outcome.  
\par In the modern formulation of quantum measurement based on
``instruments" by Davies and Lewis\cite{davieslewis}, repeatable
measurements are just a special type of measurements, and generally
the state change after the measurement---the so-called
``state-reduction"---is not presupposed. However, for continuous
spectrum observables---such as position and momentum---no projection
postulate can apply, since the eigenvectors are not normalizable,
whence they do not correspond to any physical state (in their place
the notion of ``posterior states" determined by the instrument was
introduced by Ozawa\cite{ozawacont}).  As conjectured by Davies and
Lewis\cite{davieslewis} and then proved by Ozawa in full
generality\cite{ozawacont,ozawadiscr}, for continuous spectrum no
instrument can satisfy a repeatability hypothesis, even in its weakest
conceivable form.
\par In the above scenario the orthogonal projection generally
remained a synonymous of repeatability\cite{noteOz}: however, as we
will show here, repeatable measurements are not necessarily
associated to orthogonal projectors. In the following we will completely
characterize all non orthogonal quantum measurements which are
perfectly repeatable, also providing explicit examples. We will then
show that, due to their particular structure, non orthogonal repeatable
measurements somehow ``memorize" on the system how many times the
measurement has been performed.

Due to the mentioned impossibility theorem for continuous
spectrum\cite{ozawacont,ozawadiscr}, we will consider a measurement
with discrete sample space $\Klm=\{1,2,3,\ldots\}$ as a denumerable
collection of compatible elementary events, hereafter referred to as
``outcomes". For our purpose we can also restrict the attention to the
case of pure measurements, i. e. which keep an input pure state as
pure: the generalization to mixing measurement is straightforward.

A pure measurement with discrete sample space $\Klm$ on a quantum
system is fully described by a set of contractions $\{M_e\}$ on the
Hilbert space $\sH$ of the system for each measurement outcome
$e\in\Klm$ (``contraction" means that the operator norm is bounded as
$\n{M_e}\le 1$. We remind that the squared norm $\n{A}^2$ of an
operator $A$ is defined as the supremum of $\<\psi|A^\dag A|\psi\>$
over all normalized vectors $|\psi\>\in\sH$). The state after the
measurement with outcome $e$ is given by
\begin{equation}
|\psi\>\mapsto|\psi\>_e=\frac{M_e|\psi\>}{\N{M_e|\psi\>}}\,,\label{state-red}
\end{equation}
and occurs with probability given by the Born rule
\begin{equation}
p(e)=\N{M_e|\psi\>}^2\,.
\end{equation}
Normalization of probabilities implies the completeness
\begin{equation}
\sum_{e\in\Klm} M_e^\dag M_e=I\,.
\label{eq:norm}
\end{equation}
\par We now want to determine the most general conditions under which
the measurement is perfectly repeatable. This means that the
conditional probability $p(f|e)$ of obtaining the outcome $f$ at a
repetition of the measurement, given the previous outcome was $e$,
is the Kronecker delta $p(f|e)=\delta_{ef}$. In simple words, once any
outcome is obtained, all repetitions will give the same result. In
terms of the state-reduction (\ref{state-red}), we have 
\begin{equation}
p(f|e)=\frac{\N{M_f
    M_e|\psi\>}^2}{\N{M_e|\psi\>}^2}=\delta_{ef}\qquad\forall|\psi\>\in\sH ,\;\forall e,f\in\Klm\,,
\label{eq:repeat}
\end{equation}
and, in particular, for $e=f$, Eq. (\ref{eq:repeat}) simplifies as
\begin{equation}
\N{M_e^2|\psi\>}=\N{M_e|\psi\>}\,.\label{eq:repeatsame}
\end{equation}\par 
We will now prove three lemmas, which provide a thorough mathematical
characterization of repeatable measurements, and will be helpful in
reconstructing the general form of the measurement contractions.  The
reader who is not interested in the mathematical treatment and is
seeking an intuitive understanding can jump directly to the examples
in Eqs. (\ref{eq:ex1}) and (\ref{binary}), and check himself the
repeatability condition (\ref{eq:repeat}). For the reader who is also
interested in the mathematics, only the basic theory of operators on
Hilbert spaces will be needed.\par Before stating the lemmas we will
introduce some notation. The symbol $\Ker(O)$ will denote the {\em
kernel} of the operator $O$, namely the space of all vectors on which
$O$ is null. The symbol $\Supp(O)$ denotes the {\em support} of $O$,
i. e. the orthogonal complement of the kernel, which by definition is
a subspace.  Finally, $\Rng(O)$ denotes the {\em range} of $O$,
i.e. the space of all output vectors $|\phi\>=O|\psi\>$ for any
$|\psi\>$ in the Hilbert space $\sH$. Since any contraction is bounded
and defined on all $\sH$, its kernel and range are both closed
subspaces of $\sH$, whence in the following we will use their
respective symbols to denote their closures. Also we will use the
symbol $P_\sK$ to denote the orthogonal projector on a subspace
$\sK\subseteq\sH$.
\begin{lemma} With the normalization condition (\ref{eq:norm}), the
  repeatability condition (\ref{eq:repeat}) is equivalent to
\begin{equation}\label{identity}
\left.M_e^\dag M_e\right|_{\Rng(M_e)}\equiv
P_{\Rng(M_e)}\,.
\end{equation}
Moreover, one has $M_fM_e=0$ for $e\neq f$.
\end{lemma}
\Proof That repeatability implies Eq. (\ref{identity}) follows from
identity (\ref{eq:repeatsame}). In fact, by posing
$|\varphi\>=M_e|\psi\>$ one has $\N{M_e|\varphi\>}^2=\N{|\varphi\>}^2$
for any $|\varphi\>\in\Rng(M_e)$, which implies that $M^\dag_eM_e$ is the
identity when restricted to $\Rng(M_e)$. To prove the converse
implication, we first see that Eq.  (\ref{identity}) implies that
$\N{M_e^2|\psi\>}^2=\N{M_e|\psi\>}^2$ for $|\psi\>\in\Rng(M_e)$. Then,
by applying the normalization condition (\ref{eq:norm}) and identity
(\ref{identity}) one has
\begin{equation}
M^\dag_e M_e|\psi\>=|\psi\>\equiv\sum_f M^\dag_f M_f|\psi\>\,,\label{clever}
\end{equation}
which implies that $\sum_{f\neq e}M^\dag_f M_f|\psi\>=0$, and since
the operators $M_f^\dag M_f$ are all positive, one has $M_f^\dag
M_f|\psi\>=0$ $\forall f\neq e$, then the only possibility is that
$|\psi\>\in\Ker(M_f)$ for all $f\neq e$ (due to the inclusion
$\Rng(O)\subseteq\Ker(O^\dag)^\perp$ which holds for any operator
$O$). Therefore, one has $M_fM_e|\varphi\>=0$ for all
$|\varphi\>\in\sH $.\qed
\par
An equivalent lemma is the following
\begin{lemma} With the normalization condition (\ref{eq:norm}), the
  repeatability condition (\ref{eq:repeat}) is equivalent to
\begin{equation}
\Rng(M_e)\subseteq\Ker(M_f)\,,
\label{eq:kerrng}
\end{equation}
for all $f\neq e$.
\end{lemma}
\Proof That repeatability implies Eq. \eqref{eq:kerrng} is an
immediate consequence of the previous lemma. To prove the converse
statement, consider a vector $|\psi\>\in\Rng(M_e)$. Now, Eqs.
\eqref{eq:norm} and \eqref{eq:kerrng} imply Eq. (\ref{clever}). This
means that $M^\dag_eM_e$ acts as the identity on $\Rng(M_e)$, namely
Eq. (\ref{identity}), which according to the previous lemma is
equivalent to repeatability.\qed
\par Finally we have a necessary but not sufficient condition expressed by
the following lemma.
\begin{lemma}
  With the normalization condition (\ref{eq:norm}), the repeatability
  condition (\ref{eq:repeat}) implies that $\forall e,f\in{\mathscr
    X}$, $e\neq f$
\begin{equation}\label{eq:cond3}
{\Rng(M_e)}\subseteq\Supp(M_e),\quad
{\Rng(M_e)}\perp{\Rng(M_f)}\;.
\end{equation}
\end{lemma}
\Proof We can decompose the Hilbert space $\sH $ as a direct sum
\begin{equation}
\sH =\Ker(M_e)\oplus\Supp(M_e)
\end{equation}
for all $e\in{\mathscr X}$. Now suppose by absurdum that a vector
$|\psi\>\in\sH $ exists such that
\begin{equation}
M_e|\psi\>=|v\>+|\psi^\prime\>\,,
\end{equation}
with $|v\>\in\Ker(M_e)$ and $|\psi^\prime\>\in\Supp(M_e)$.
Then, since $\n{M_e}\le 1$, using Eq. (\ref{eq:repeatsame}) we have
\begin{equation}
\N{\psi^\prime}^2\geq\N{M_e\psi^\prime}^2=\N{M_e^2\psi}^2=\N{M_e\psi}^2=\N{v}^2+\N{\psi^\prime}^2\,,
\end{equation}
and this is possible if and only if $|v\>=0$. Therefore, we have
$\Rng(M_e)\subseteq\Supp(M_e)$. This relation along with Eq.
\eqref{eq:kerrng} gives the orthogonality between the closures of the
ranges, since
${\Rng(M_e)}\subseteq\Ker(M_f)=\Supp(M_f)^\perp\subseteq{\Rng(M_f)}^\perp$.\qed
\par From the last lemma it follows that only for finite dimensional
$\sH $ we have the customary orthogonal measurement paradigm.
\begin{corollary}
For finite dimensional $\sH $ a measurement is repeatable iff it is orthogonal.
\end{corollary}
\Proof
For finite dimensional $\sH $ the support and the range of any
operator have the same dimension, and this fact along with the first
condition in Eq.  (\ref{eq:cond3}) implies
$\Rng(M_e)\equiv\Supp(M_e)$.  Thus the operators $M^\dag_e M_e=P_e$,
for $e\in\Klm$, form an orthogonal projective POVM, namely
\begin{equation}
P_eP_f=\delta_{ef}P_f\,.
\label{eq:findim}
\end{equation}
\par For infinite dimensional $\sH $, on the contrary, we cannot draw
the same conclusion, since a subspace can have the same (infinite)
dimension of a space in which it is strictly included. And, in fact,
it is easy to construct counterexamples of repeatable measurements, as
that given in the following, which satisfy conditions \eqref{identity}
or (\ref{eq:kerrng}), and do not satisfy the stronger orthogonality
condition (\ref{eq:findim}).
\medskip\par\noindent{\bf Example. }{\em 
The following set of contractions
\begin{equation}
M_l=\sqrt{p_l}|l\>\<0|+\sum_{j=0}^\infty|n(j+1)+l\>\langle
nj+l|,\;1\le l\le n
\label{eq:ex1}
\end{equation}
with $p_l\geq0$, $\sum_lp_l=1$ and $|n\>$ a generic
discrete basis for the Hilbert space, defines a perfectly repeatable pure
measurement with sample space $\Klm=\{1,2,\ldots,n\}$. }\medskip
\par\noindent
That the set of operators $\{M_e\}$ in Eqs. \eqref{eq:ex1} actually describes a
measurement follows by just checking the normalization
(\ref{eq:norm}). Moreover the set of operators satisfies
condition \eqref{identity}, as well as condition \eqref{eq:kerrng},
whence they describe a repeatable measurement. On the other hand, the
measurement is not orthogonal, since the corresponding
POVM is given by 
\begin{equation}
P_l=p_l|0\>\<0|+\sum_{j=0}^\infty|nj+l\>\langle nj+l|,\;1\le l\le n.
\end{equation}
We emphasize that the same POVM also describes a non repeatable
measurement, such as that corresponding to the set of contractions
\begin{equation}
N_l=\sqrt{p_l}|0\>\<0|+\sum_{j=0}^\infty|nj+l\>\langle nj+l|,\;1\le l\le n.
\end{equation}
This fact evidences that repeatability is a feature which is obviously
related to the state-reduction of the measurement, not to the POVM,
\eg one can have an orthogonal POVM for a non repeatable measurement.
\par
\begin{figure}[h]
\epsfig{file=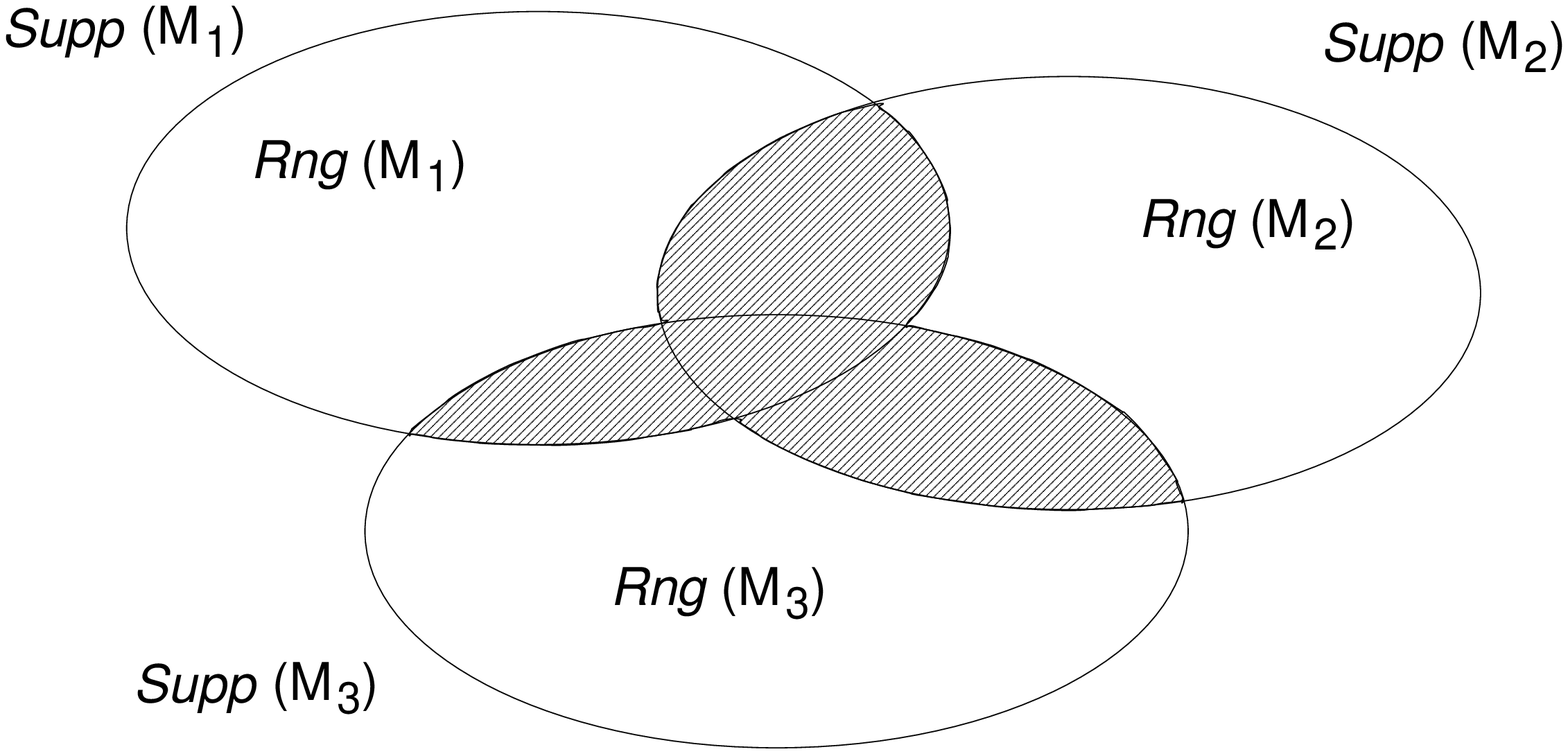,width=6cm}
\caption{Illustration of the relations between supports and ranges
of the contractions of a generic repeatable non orthogonal measurement
with three outcomes.} \label{fig1}
\end{figure}

 At this point, the question is how to characterize a generic
non orthogonal repeatable measurement, namely which is the general
form of the contractions $\{M_e\}$ that satisfy Eq.  \eqref{identity}
or Eq.  \eqref{eq:kerrng}. The necessary conditions \eqref{eq:cond3}
now come at hand: if we exclude the case of orthogonal measurements,
then there must exist at least one $M_i$ such that one has the strict
inclusion ${\Rng(M_i)}\subset\Supp(M_i)$.  We can now decompose the
subspace $\Supp(M_i)$ in orthogonal components as follows
\begin{equation}
\Supp(M_i)=\Rng(M_i)\oplus{\mathcal C}(M_i)\,, 
\end{equation}
where ${\mathcal C}(M_i)$ is the orthogonal complement of $\Rng(M_i)$
in $\Supp(M_i)$. The operator $M_i$ on its support can then be written
as
\begin{equation}
M_i=V_i+W_i\,,\label{eq:sepm}
\end{equation}
with $\Supp(V_i)=\Rng(M_i)$ and $\Supp(W_i)={\mathcal
  C}(M_i)$. 
The normalization of the POVM implies
\begin{equation}
\sum_e(V^\dag_eV_e+W^\dag_eW_e+V^\dag_eW_e+W^\dag_eV_e)=I\,.
\end{equation}
Since they represent off-diagonal operators, the cross terms must be
null. More precisely, one must have $W^\dag_eV_e=0$ for each term
separately, since due to orthogonality of supports for different $e$
these terms are all linearly independent, and similarly
$V^\dag_eW_e=0$ by orthogonality of ranges.
These facts along with Eq. \eqref{identity}---which states that $M_i$
is isometric on its range---implies that $V_i$ is a partial isometry
\begin{equation}
V^\dag_i V_i=P_i\equiv P_{{\Rng(M_i)}}\,,
\end{equation}
and we can rewrite the normalization condition \eqref{eq:norm} as
\begin{equation}
\sum_e P_e+\sum_fW^\dag_fW_f=I\,.
\end{equation}
The only conditions that the operators $W_e$ must obey are then 
\begin{equation}
\begin{split}
  &\Supp(W_e)={\mathcal C}(M_e)\;,\\
  &{\Rng(W_e)}\subseteq{\Rng(M_e)}\quad\Rightarrow\quad W^\dag_fW_e=0\,,\ e\neq f\;,\\
  &\sum_e W^\dag_e W_e=P_\sK\;,
\end{split}
\end{equation}
where $P_\sK$ is the projection on the intersection space
$\sK=\left[\bigoplus_e\Rng(M_e)\right]^\perp$. For some events $f$ the
operator $W_f$ could be null, namely $\Supp(M_f)\equiv{\Rng(M_f)}$:
when this holds for all events $f\in\mathscr{X}$, the described
measurement is just the conventional orthogonal one.  Summarizing, for
a non orthogonal repeatable measurement, the contractions $M_e$ have
supports that intersect, at least for a couple of events $e$, but
their ranges fall outside the intersection, as represented in Fig.
\ref{fig1}.
They act as an isometry on their ranges, while on the intersection
space $\sK$ it is the sum of $W_e^\dag W_e$ that acts as the
identity.  Notice that each operator $W_e$ needs not to be
singularly proportional to a partial isometry, as in the example
given before. In fact, consider the binary measurement described by
the contractions 
\begin{equation}
\begin{split}
  M_1=&\sqrt{p_1}|2\>\<0|+\sqrt{p_2}|4\>\langle1|+\sum_{n=1}^\infty|2(n+2)\>\langle2n|\\
  M_2=&\sqrt{1-p_1}|3\>\<0|+\sqrt{1-p_2}|5\>\langle1|\\
  &+\sum_{n=1}^\infty|2(n+2)+1\>\langle2n+1|\,;\label{binary}
\end{split}
\end{equation}
the corresponding POVM is given by
\begin{equation}
\begin{split}
  P_1=&p_1|0\>\<0|+p_2|1\>\langle1|+\sum_{n=1}^\infty|2n\>\langle2n|\\
  P_2=&(1-p_1)|0\>\<0|+(1-p_2)|1\>\langle1|\\
  &+\sum_{n=1}^\infty|2n+1\>\langle2n+1|\,.
\end{split}
\end{equation}
In this case
$W_1=\sqrt{p_1}|2\>\<0|+\sqrt{p_2}|4\>\langle1|$ and
$W_2=\sqrt{1-p_1}|3\>\<0|+\sqrt{1-p_2}|5\>\langle1|$,
and $W^\dag_iW_i$ are not proportional to orthogonal
projectors, since
\begin{equation}
\begin{split}
&W^\dag_1W_1=p_1|0\>\langle 0|+ p_2|1\>\langle 1|\\
&W^\dag_2W_2=(1-p_1)|0\>\langle 0|+ (1-p_2)|1\>\langle 1|\;,
\end{split}
\end{equation}
while, clearly,
\begin{equation}
W^\dag_1W_1+W^\dag_2W_2=|0\>\langle 0|+|1\>\langle 1|=P_\sK\;.
\end{equation}
\par We are now in position to state the general form of a POVM
$\{P_e\}$ admitting a repeatable measurement. One must have
\begin{equation}
\begin{split}
  &P_e=Z_e+T_e\,,\quad e\in\Klm\,,\\
  &Z_e T_f=T_f Z_e=0\,,\quad\forall e,f\in\Klm\,,\\
  &T_e\geq0\,,\quad\sum_{e\in\Klm}T_e=Z_\omega\,,\\
  &Z_eZ_f=Z_e\ \delta_{ef}\,,\quad \forall e,f\in\Klm\cup\{\omega \}\,,\\
\end{split}
\end{equation}
with the normalization $\sum_{e\in{\mathscr X}}P_e\equiv
Z_{\omega}+\sum_{e\in{\mathscr X}}Z_i=I$.
The orthogonal case corresponds to $T_e=0$, $\forall e\in\Klm$.\par
Let's now see how a ``memory" of the number of performed repetitions
is associated to a non orthogonal repeatable measurement.  This is a
consequence of a theorem by Wold and von Neumann\cite{Murphy,halmos}
which states that every isometry can be written as a direct sum of
unilateral shift operators and possibly a unitary (an unilateral shift
$S$ can always be written in the form
$S=\sum_{j=1}^\infty|j+k\>\langle j|$, $k\ge 1$, for a suitable
orthonormal basis $\{|j\>\}$). The operators $V_e$ in Eq.
\eqref{eq:sepm} can then be further separated in the direct sum
$V_e=U_e+S_e$ of a unitary $U_e$ and a pure isometry $S_e$, and we
have
\begin{equation}
M_e=V_e+W_e=U_e+S_e+W_e\,.
\end{equation}
Let us now consider an initial state $|\psi\>$ with non-vanishing
component in the support of $M_e$, and suppose that the outcome $e$
occurred. Since $V^\dag_eW_e=0$, one can equivalently write
$S^\dag_eW_e=0$ and $U^\dag_eW_e=0$. The latter identity implies that
the range of $W_e$ is orthogonal to $\Supp(U_e^\dag)\equiv\Supp(U_e)$,
and thus the conditional state
$|\psi_e\>=\frac{M_e|\psi\>}{\n{M_e|\psi\>}}$ cannot be in the support
of $U_e$, namely it must belong to the support of $S_e$. Therefore,
for the successive measurements we will effectively have $M_e\equiv
S_e$, and successive applications will shift the observable $\{|j\>\<
j|\}$ to $\{|j-k\>\< j-k|\}$, where $\{|j\>\}$ is the orthonormal
shifted basis for any chosen unilateral shift component of $S_e$.
Notice that the index $j$ can be checked without affecting the
repeatability of the outcome $e$.
\par In summary, we have shown that there exist non orthogonal perfectly
repeatable measurements, and only for finite dimensions repeatability
is equivalent to orthogonality. On the contrary, for infinite
dimension there exist non orthogonal repeatable measurements, of which
we have given the most general form, based on necessary and sufficient
conditions, and providing some explicit examples. Finally, we have
shown how the measured system undergoing such a measurement must
retain some ``memory" of the number of times that the measurement was performed.   
\par G. M. D. acknowledges discussions with M. Ozawa supporting the
present general structure of non orthogonal repeatable
measurements. This work has been 
sponsored by INFM through the project PRA-2002-CLON, and by EEC and
MIUR through the cosponsored ATESIT  
project IST-2000-29681 and Cofinanziamento 2002. 
G. M. D. also acknowledges partial support from the MURI
program administered by the Army Research Office under Grant 
No. DAAD19-00-1-0177.


\begin{thebibliography}{99}
\bibitem{Mittelstaedt} P. Mittelstaedt, {\em The interpretation of
    Quantum Mechanics and the Measurement Process} (Cambridge
  University Press, Cambridge, 1998).
\bibitem{vonneu} J. Von Neumann, \emph{Mathematical principles of
    Quantum Mechanics} (Princeton University Press, 1955).
\bibitem{Lud51} G. L\H{u}ders, Ann. Physik {\bf 8} (6), 322 (1951).
\bibitem{davieslewis} E. B. Davies and J. T. Lewis, Commun. Math. Phys. {\bf 17}, 239 (1970).
\bibitem{ozawacont} M. Ozawa, J. Math. Phys. {\bf 25}, 79 (1984).
\bibitem{ozawadiscr} M. Ozawa, Publ. Res. Inst. Math. Sci., Kyoto Univ.
  {\bf 21}, 279 (1985).
\bibitem{noteOz} The fact that orthogonal projection follows from
further assumptions in addition to repeatability was emphasized by
Ozawa in many papers: see for example Ref. \cite{ozawanondeg}.
\bibitem{ozawanondeg} M. Ozawa, Phys. Rev. A {\bf 62}, 062101 (2000).
\bibitem{Murphy} G. J. Murphy, {\em $C^*$-Algebras and Operator
    Theory} (Academic Press, London, San Diego, 1990).
\bibitem{halmos} P. R. Halmos, \emph{A Hilbert space problem book}
  (American Book, New York, 1967).
\end{thebibliography}
\end{document}